# Structure and Magnetic Properties of Vacuum-Annealed CoFeB Thin Films: From Amorphous Alloy to Metastable $(Co,Fe)_{23}B_6$ τ-Boride

*Petr Shvets*[*a], *Grigorii Kirichuk*[a], *Vitalii Salnikov*[b], *Jacques O'Connell*[c], *Valeria Rodionova*[b], *Alexandr Goikhman*[d], *Ksenia Maksimova*[a]

E-mail: pshvets@kantiana.ru

[a]Research and Educational Center "Functional Nanomaterials", Immanuel Kant Baltic Federal University, Kaliningrad, Russian Federation

[b]Research and Educational Center "Smart Materials and Biomedical Applications", Immanuel Kant Baltic Federal University, Kaliningrad, Russian Federation

[c]Nelson Mandela University, Port Elizabeth, South Africa

[d]Koenigssystems UG, Schenefeld bei Hamburg, Germany

Controlled crystallization of amorphous alloys offers a powerful route to tailor magnetic and structural properties at the nanoscale. Thin films of CoFeB alloy are essential for the development of various spintronic devices. The crystallization mechanisms of CoFeB during the annealing process have been thoroughly investigated in earlier studies, demonstrating that boron diffuses from the amorphous film, allowing the remaining CoFe to form a body-centered cubic lattice. Here, a distinct transformation pathway in pulsed-laser-deposited amorphous $Co_{40}Fe_{40}B_{20}$ films is revealed, where vacuum annealing drives the formation of a metastable τ-boride phase, $(Co,Fe)_{23}B_6$. Comprehensive structural characterization – combining X-ray diffraction, transmission electron microscopy, and compositional analysis – proves that τ-boride forms with high crystalline quality and minimal boron loss. Following the transition from amorphous CoFeB films to crystalline $(Co,Fe)_{23}B_6$, an improvement of magnetic properties is observed, with corresponding increases in such values as saturation magnetization, coercivity, loop squareness, and average magnetic moment. The reproducible stabilization of a boron-rich metastable phase in CoFeB thin films expands the known crystallization landscape of this technologically important alloy system. These findings provide new insight into phase engineering in transition-metal borides and open opportunities for designing nanostructured magnetic materials with tunable functionality for next-generation spintronic and nanoelectronic devices.

## 1. Introduction

Thin films composed of the ferromagnetic alloy CoFeB have become an essential functional layer for spintronic applications, including spin-orbit torque and spin-transfer torque magnetic random-access memory devices,[1,2] spin currents generators operating on the spin Hall effect,[3,4] nano-oscillators[5] and systems designed for neuromorphic computing.[6] One of the primary research focuses in CoFeB-based structures is the formation of perpendicular magnetic anisotropy (PMA) in heavy metal/CoFeB/MgO multilayers.[7,8] This unique property enables the fabrication of magnetic tunnel junctions (MTJs), which are critical for next-generation magnetic memory technologies.[9]

To improve the functional properties of MTJ structures, thermal annealing is commonly applied.[10,11] This process promotes the formation of PMA and increases the interface anisotropy energy. Additionally, it triggers the crystallization of CoFeB, which have been thoroughly investigated through various techniques.[12-21] These studies demonstrate that at elevated temperatures, boron migrates from the amorphous CoFeB matrix to the interfaces or into the neighboring layers, while the remaining CoFe crystallizes into a body-centered cubic structure. CoFe exhibits a (100) texture when crystallization starts at the MgO/CoFeB interface, whereas a (110) texture suggests that crystallization initiates from the interface with a metal such as Ru or Ta. At higher annealing temperatures, the formation of boride phases, $(Co,Fe)_3B$, is sometimes reported.[14,16]

In our study, we aimed to investigate the crystallization process of the films produced using pulsed laser deposition from a standard $Co_{40}Fe_{40}B_{20}$ target. In line with previous reports, we first obtained amorphous samples at room temperature and subsequently annealed them under high-vacuum conditions. Unexpectedly, most of the resulting films crystallized into the high-quality metastable $(Co,Fe)_{23}B_6$ phase. This phase belongs to the family of τ-borides, with the structure of the $Cr_{23}C_6$ type.[22] This structure has a complex face-centered cubic lattice (space group $Fm\bar{3}m$) that includes four unique metal sites and one site for carbon (boron). The unit cell contains 116 atoms and has a lattice constant of approximately 1 nm. τ-borides are metastable,[23] and their synthesis is challenging, especially in the form of the single-phase samples.[24-28] Additionally, there is a significant lack of information regarding the formation of τ-borides as nanostructured thin films,[29-31] leaving their properties largely unexplored. In the current study, we seek to address this gap by systematically analyzing the crystallization process of $(Co,Fe)_{23}B_6$ in relation to the annealing conditions and providing the results regarding the magnetic properties of the films. Our work enhances the understanding of the crystallization mechanisms and the properties of the material, which is essential for the development of spintronics.

## 2. Results and discussion

### 2.1. Elemental composition

The notation "CoFeB" or "FeCoB" usually refers to the composition of $(Co,Fe)_{80}B_{20}$, where the atomic concentrations of Co and and Fe are typically equal. During the sputtering process of thin films, boron may be transferred from the target in a non-stoichiometric manner. However, it is quite challenging to detect and quantify light elements such as boron in thin films, which often leaves the precise composition of the film unknown. Potential solutions to this issue require the application of sophisticated techniques like nuclear reaction analysis. For example, in the case of an $Fe_{52}Co_{28}B_{20}$ target, the film composition obtained through magnetron sputtering was found to be $Fe_{54.8}Co_{30}B_{15.2}$, indicating a notable decrease in boron content.[32]

CoFeB films are typically deposited using magnetron sputtering, so a reduction of boron concentration in the film relative to the target can be expected in many previously published studies. In our investigation, we employed pulsed laser deposition (PLD), a method recognized for its capability to facilitate the stoichiometric transfer of material from a target to a film.[33] To confirm this for the CoFeB target, we created a thick film (~200 nm) and analyzed its elemental composition through Rutherford backscattering spectrometry (RBS) and energy dispersive X-ray spectroscopy (EDX). The intensity of the RBS signal in the Co/Fe region (channels 260-360, **Figure 1a**) is affected by the boron concentration in the film, while the signal intensity from the substrate (channels 170 and below) remains unchanged. The best fit indicates a boron concentration of approximately 20 atomic percent. This stoichiometric transfer of boron is atypical for thin film deposition and may be a crucial factor in the formation of the metastable $(Co,Fe)_{23}B_6$ phase, which demands a considerable amount of boron for crystallization (20.7%).

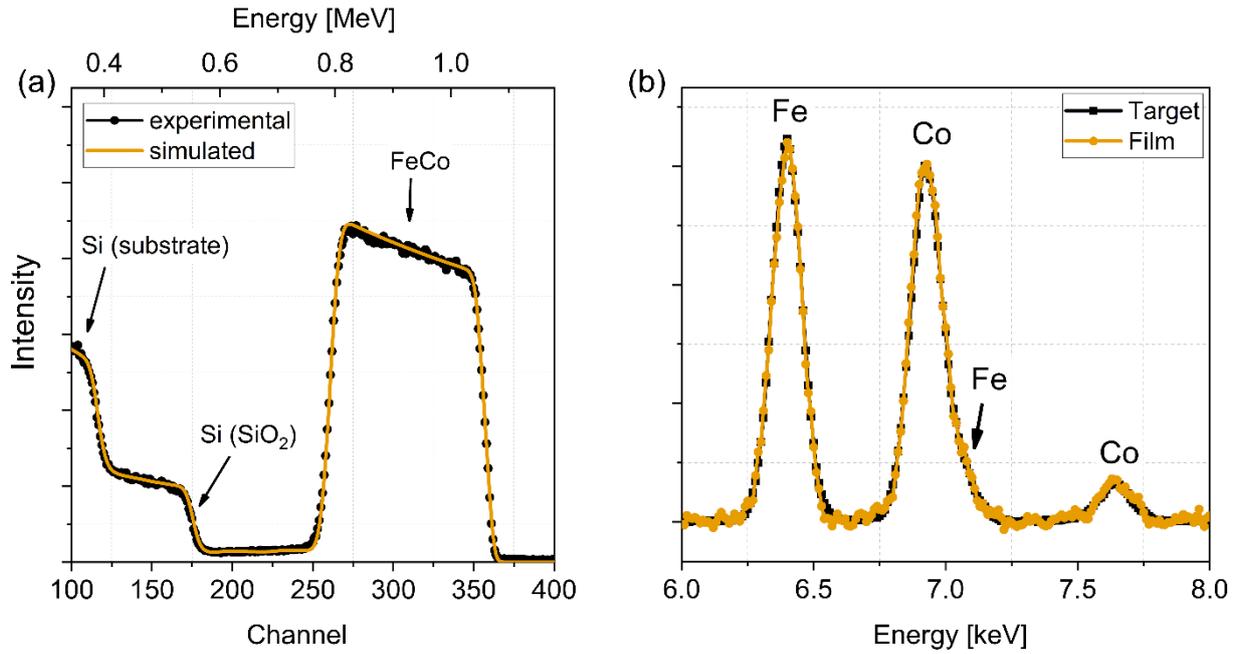

**Figure 1.** (a) The RBS spectrum for CoFeB films deposited on a Si/SiO$_2$ substrate. (b) EDX spectra for the CoFeB target and the CoFeB film deposited using PLD.

Given that the atomic masses of cobalt and iron are nearly identical, RBS cannot be utilized for an accurate evaluation of the Fe to Co ratio. Therefore, EDX was used to determine this ratio, revealing a composition of roughly 46 to 54 atomic percent in both the target material and the film (Figure 1b).

**2.2. Crystal structure**

In the following discussion, we will focus on thin films with a thickness of less than 30 nm that have undergone annealing under different conditions. In our research, we varied the annealing temperature, cooling rate, and film thickness; comprehensive information about these conditions and sample designations can be found in **Table S1.1**.

*2.2.1. XRD*

The X-ray diffraction (XRD) pattern of a typical thin $(Co,Fe)_{23}B_6$ film is primarily characterized by the (333) reflection at $2\theta \sim 45°$. In low-quality samples, this may be the sole peak observable in Bragg-Brentano geometry, which can lead to confusion with the most intense reflections of CoFe [(110), space group $Pm\bar{3}m$] or $(Co,Fe)_2B$ [(211), space group $I4/mcm$]. For high-quality films, a series of six (nnn) (n = 1 … 6) reflections can be distinguished, as shown in **Figure 2a**. Besides $M_{23}B_6$, this set of peaks may also be characteristic of $(Co,Fe)_3B$ [(nn0), n = 1 … 6, space group $I\bar{4}$], necessitating further phase confirmation. Additional peaks of $M_{23}B_6$ were observed in various geometries (refer to Figure 2b,c and **Table S2.1**). All observed peaks align perfectly with the theoretical powder pattern of $M_{23}B_6$ (Figure 2d), which was computed using Profex software,[34] positional parameters from literature,[26] and a lattice constant of $a = 10.52$ Å.

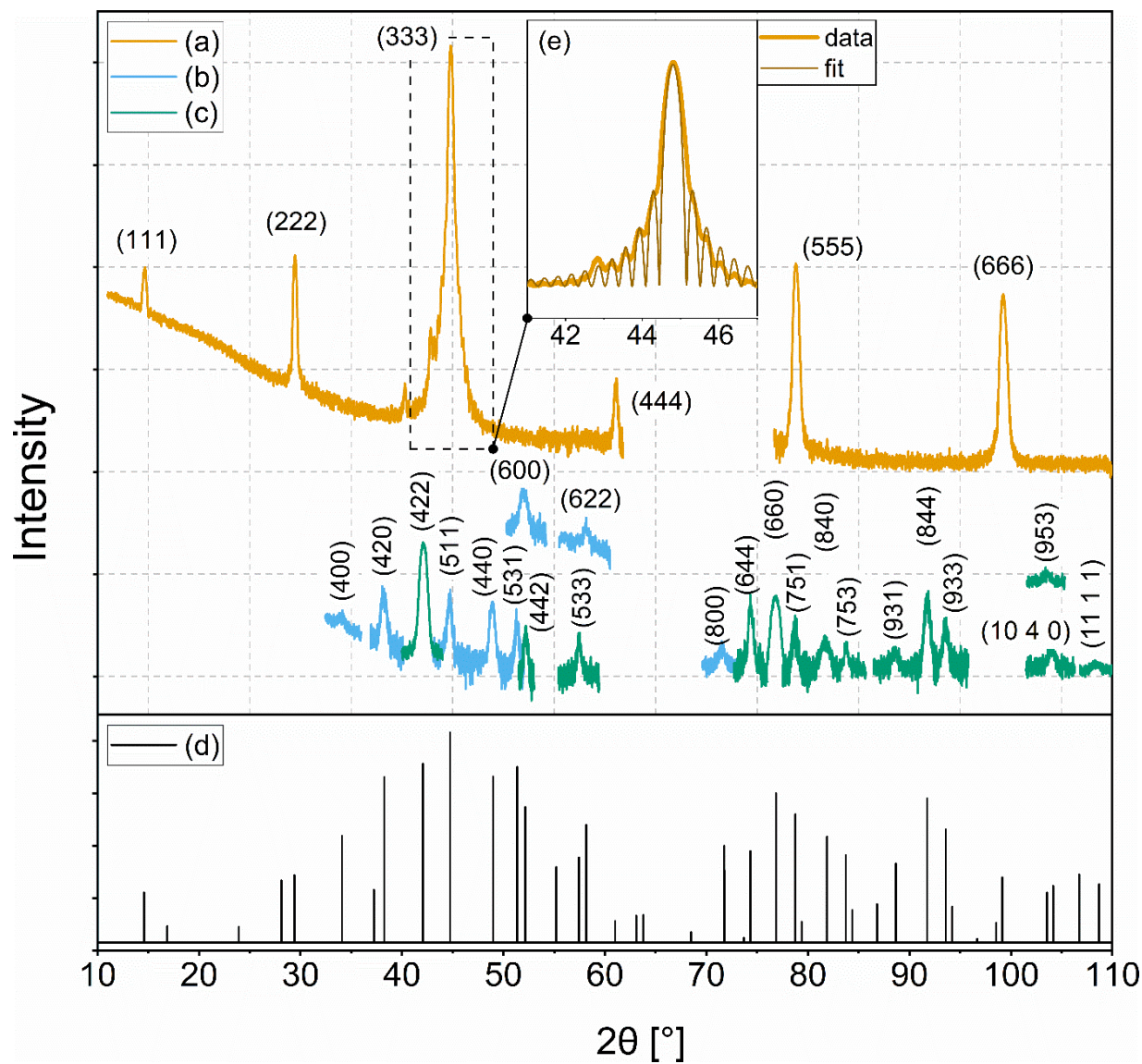

**Figure 2.** XRD patterns for sample STh-4: (a) θ–2θ geometry, (b) tilted sample geometry, (c) asymmetrical geometry. (d) Theoretical XRD pattern of $M_{23}B_6$ powder. (e) The section of the (333) peak illustrating Laue oscillations. All graphs are presented in a logarithmic scale.

The presence of Laue oscillations (Figure 2e) indicates the high quality of the produced $(Co,Fe)_{23}B_6$ film. The fitting curve corresponding to these oscillations was calculated using the simple Laue function:[35]

$$I(Q) \propto \frac{\sin(0.5NQc)^2}{\sin(0.5Qc)^2},$$

where $Q$ is the scattering vector ($Q = 4\pi \sin\theta / \lambda$, $\lambda = 1.5406$ Å), $c$ is the out-of-plane lattice parameter ($c = 2.021$ Å), and $N$ is the number of coherently diffracting unit cells. Through the fitting procedure, we can derive $N$ and calculate the total thickness of the coherently diffracting crystal as $c \times N$.

In our experiments, we varied the sample thickness, annealing temperature, cooling rate, and substrate type. From XRD measurements, we discovered that the $(Co,Fe)_{23}B_6$ phase was prevalent across a broad range of conditions. The formation of this boron-reach phase implies that, in our case, the diffusion of boron into the substrate is minimal, which is contrary to typical observations.[10,36,37]

As the sample thickness increased, we observed a narrowing of the $(Co,Fe)_{23}B_6$ reflexes (consistent with the Scherrer equation) and a rise in their intensity (see **Figure S2.1**). In all cases (ranging from 6 to 27 nm), the thicknesses obtained from Laue oscillations and XRR curves were identical (refer to Table S1.1), demonstrating the high quality and single-phase nature of the crystal growth.

The cooling rate might be a crucial parameter because the bulk $M_{23}B_6$ phase is typically obtained from the uncooled state and can only be preserved at room temperature at very high cooling rates (greater than around 25 °C min$^{-1}$).[27] However, for thin films, this is not essential, and perfect $(Co,Fe)_{23}B_6$ can be produced even at cooling rates as low as 2 °C min$^{-1}$ (refer to **Figure S2.2**).

The critical parameter that influences the crystal structure of the samples is the annealing temperature (**Figure 3**). For temperatures under 350 °C, the film keeps its initial amorphous state. At 400 °C, $(Co,Fe)_{23}B_6$ is produced and remains the only phase until 700 °C. At 750 °C, another phase, $(Co,Fe)_2B$ (space group I4/mcm), is produced. To confirm the presence of $(Co,Fe)_2B$, several additional reflexes were detected in different geometries; see Table S2.2 and Figure S2.3 for further details. At 800 °C, $(Co,Fe)_{23}B_6$ disappears, and the film becomes a mixture of $(Co,Fe)_2B$ and CoFe (refer to **Table S2.3** and **Figure S2.4** for confirmation of the CoFe phase).

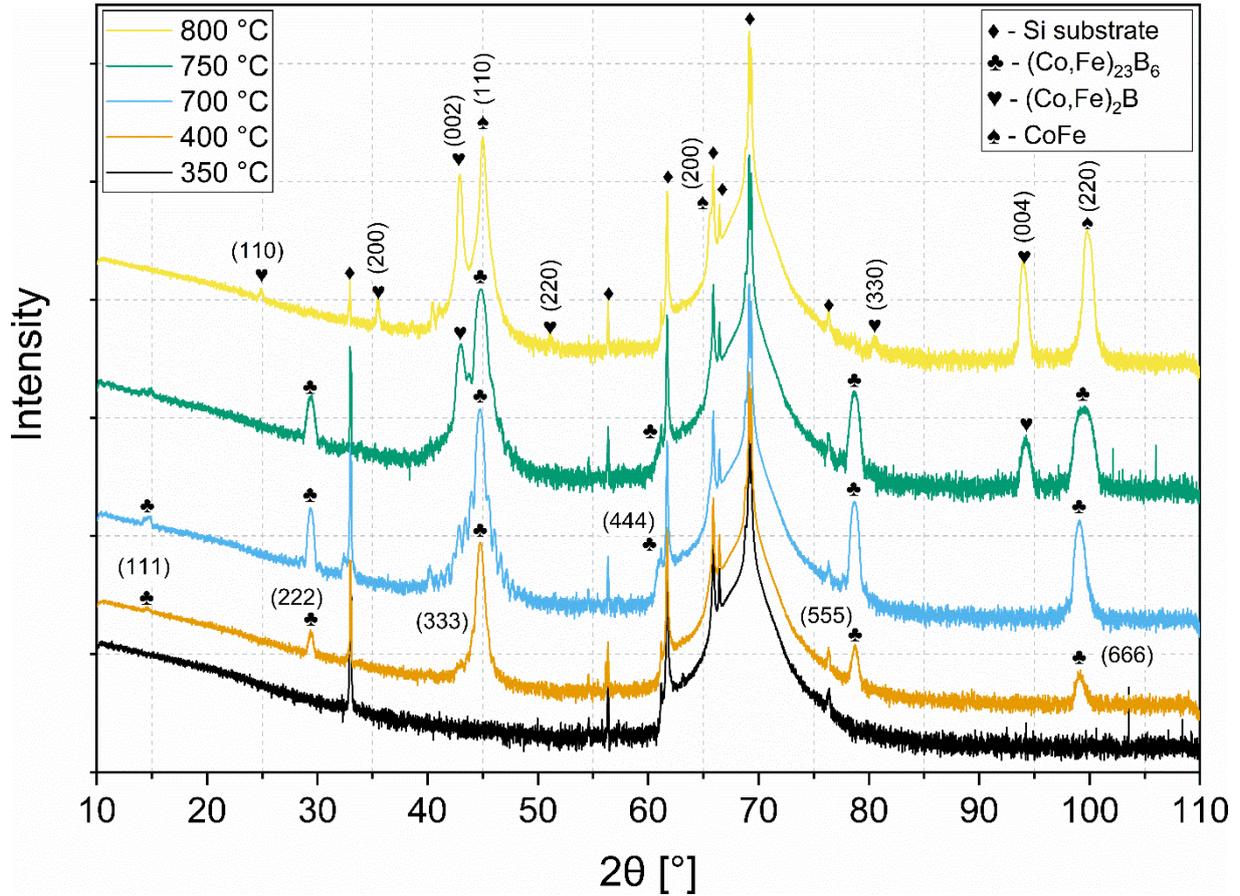

**Figure 3.** XRD patterns for CoFeB films annealed at different temperatures (samples STa-2, STa-3, STa-6 – STa-8). All graphs are presented in a logarithmic scale.

To estimate the lattice strains ($\varepsilon$) in $(Co,Fe)_{23}B_6$ samples produced at different temperatures, we applied the Williamson-Hall method[38] using the equation

$$\beta \cos\theta = K\lambda/CS + 4\varepsilon \sin\theta$$

where $\theta$ refers to the position of the XRD peak, $\beta$ is its width, $K$ is a constant ($K = 0.9$), $\lambda$ is the X-ray wavelength, and $CS$ stands for crystallite size. The resulting strains are very low (~$10^{-4}$, see **Table S2.4**), and the lattice constant is similar for all samples, with $a = 10.52$ Å. The crystallite size determined by this method aligns with the thickness of the film, as expected for high-quality samples. The surface morphology of the films was studied by atomic force microscopy (AFM); refer to section 3 of the Supporting Information for further details.

So far, our research was focused on the structural properties of CoFeB films deposited only on Si / $SiO_2$ substrates. However, as mentioned previously, the interface between CoFeB and MgO could be particularly significant. Thus, we conducted a series of experiments aimed at investigating the crystallization of CoFeB films grown on single-crystal MgO (100) substrates. At temperatures reaching 600 °C and higher, we detected the formation of (200) CoFe (**Figure S2.5**), consistent with earlier studies.[12,14,16,17] The CoFe film exhibits epitaxial growth, with its cube edges oriented at a 45° angle relative to the in-plane MgO (100) direction (see **Figure S2.6** for the corresponding φ-scans). At lower temperatures, the $(Co,Fe)_{23}B_6$ phase is formed. This phase is of high quality, exhibiting visible Laue oscillations, but it lacks any epitaxy, showing randomly distributed in-plane orientations.

*2.2.1. TEM*

**Figure 4a,b** shows bright field (BF) transmission electron microscopy (TEM) images of the Mo coated CoFeB layer. The layer thickness is approximately 12 nm, consistent with XRD and X-ray reflectivity (XRR) measurements. The diffraction contrast highlights regions with different CoFeB grain orientations, which allows for the estimation of the in-plane grain size to be a few tens of nm. In certain grains, crystal planes are visible (Figure 4c,d). In Figure 4c, these planes are parallel to the substrate, with an interplanar distance of about 6 Å. This measurement is in close agreement with the interplanar distance between (111) planes in the $(Co,Fe)_{23}B_6$ phase, which is 6.07 Å. In Figure 4d, the planes have the same interplanar distance but are rotated by 71°. This rotation corresponds to the angle between (111) and $(11\bar{1})$, so the observation of such planes in a (111)-textured film is expected. The selected area electron diffraction pattern (Figure 4e) also confirms the formation of the $(Co,Fe)_{23}B_6$ phase and its (111) texture. The most intense spots correspond to diffraction of the electron beam directed along [112] in $(Co,Fe)_{23}B_6$. There are also some additional spots corresponding to the $[2\bar{3}\bar{1}]$ direction, as confirmed by diffraction simulations (Figure 4f).

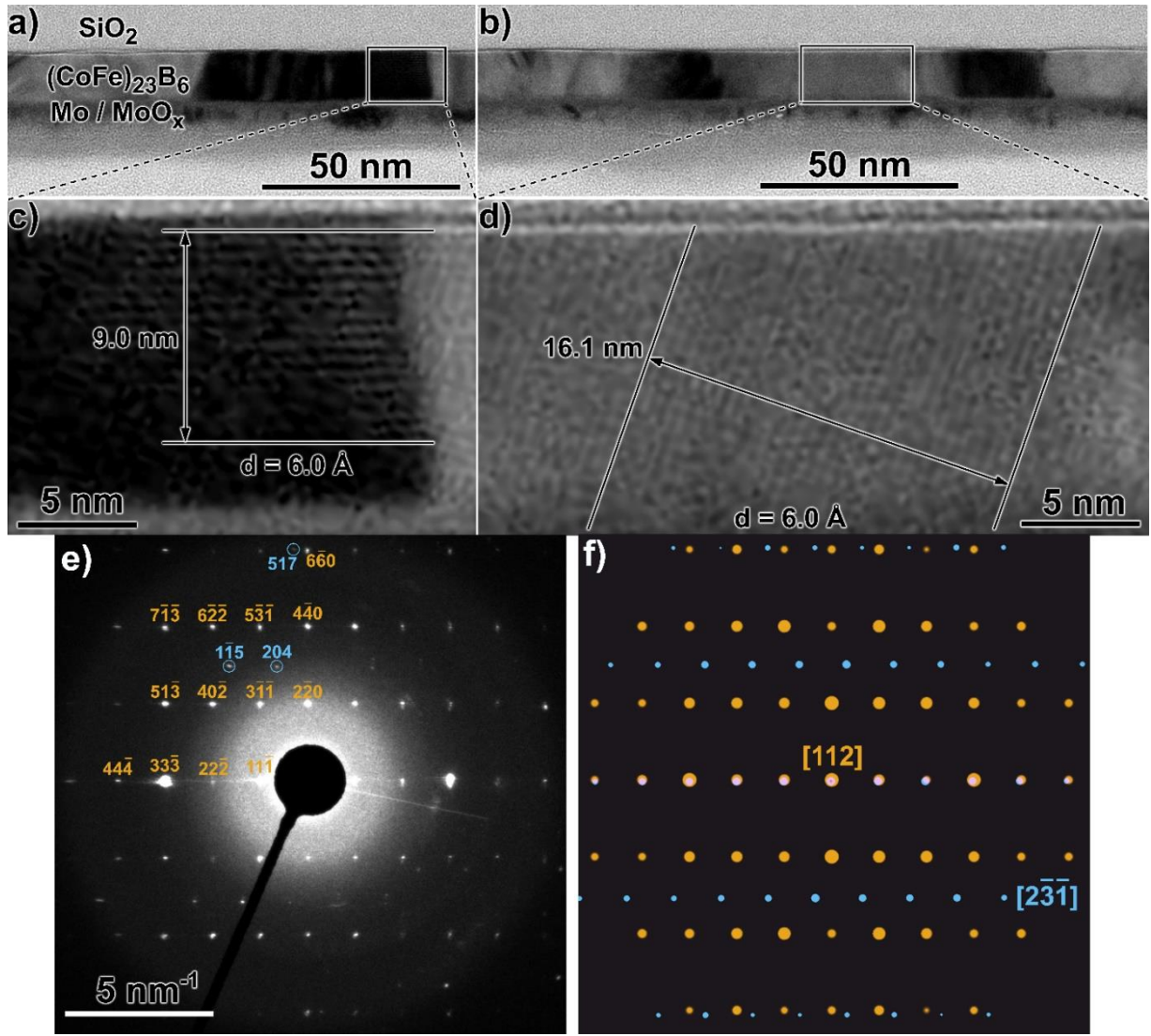

**Figure 4.** (a-d) Cross-sectional BF TEM images of the (Co,Fe)$_{23}$B$_6$ film. (e) Experimental and (f) simulated electron diffraction patterns of the film.

## 2.3. Magnetic properties

It is of particular interest to investigate the magnetic properties of CoFeB films as they crystallize into the (Co,Fe)$_{23}$B$_6$ phase. For this study, we selected a series of samples produced at various temperatures on silicon substrate, because these samples exhibit different crystal structures (for samples on MgO substrates, the results of the magnetic measurements are shown in sections 4 and 5 of the Supporting Information). The correlation between the magnetic properties of CoFeB films and their crystal structure is clear (**Figure 5**). It is important to note that the magnetic measurements described below were conducted on unprotected CoFeB, which may lead to oxidation of the film's surface. To demonstrate that the oxide would not significantly alter the magnetic properties, a series of samples with different capping layers were prepared and subsequently analyzed by vibrating sample magnetometry (VSM). The results are described in section 6 of the Supporting Information.

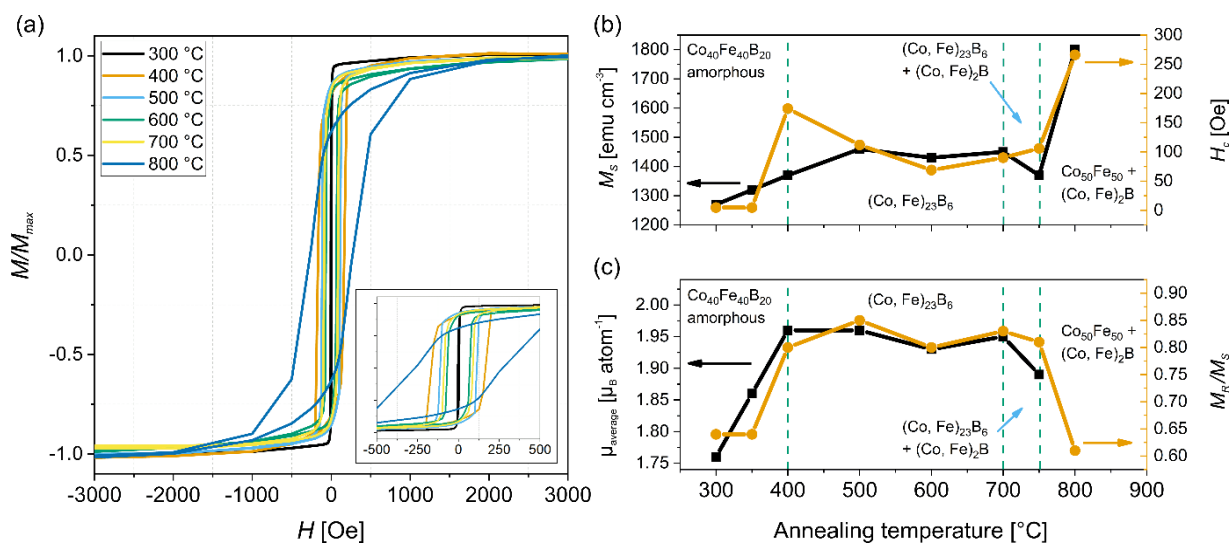

**Figure 5.** (a) Field dependence of magnetization measured at room temperature (the inset shows an enlarged view), (b) the dependence of saturation magnetization $M_S$ (the black line) and coercivity $H_C$ (the orange line) on annealing temperature, (c) the dependence of the average Co/Fe atomic magnetic moment (the black line) and the $M_R/M_S$ ratio (the orange line) on annealing temperature.

The saturation magnetization $M_S$ of the (Co,Fe)$_{23}$B$_6$ phase (400°C ≤ $T_a$ ≤ 700 °C) achieves approximately 1400–1450 emu cm$^{-3}$, while the amorphous film ($T_a$ ≤ 350 °C) exhibits a lower value of 1270 emu cm$^{-3}$. The decrease in $M_S$ at 750 °C is attributed to the formation of the (Co,Fe)$_2$B phase, which has a lower magnetization.[39,40] At 800 °C, the crystallization of CoFe grains begins, leading to a significant rise in $M_S$, since this phase has a saturation magnetization of up to 1900 emu cm$^{-3}$.[41] Such magnetic behavior is in agreement with our structural data. The experimentally obtained $M_S$ values for the (Co,Fe)$_{23}$B$_6$ phase are close to theoretical predictions:[23] 1500 emu cm$^{-3}$ for Fe$_{23}$B$_6$ and 920 emu cm$^{-3}$ for Co$_{23}$B$_6$. However, the values we obtained are somewhat higher than those previously reported: 1350 emu cm$^{-3}$ for Fe$_{23}$B$_6$[42] and ~880 emu cm$^{-3}$ for Co$_{23}$B$_6$[28]. Furthermore, the (Co,Fe)$_{23}$B$_6$ phase is of particular interest because of its elevated Co/Fe atomic magnetic moments. The method for calculating the average atomic magnetic moment is described in section 7 of the Supporting Information. It is important to note that the magnetic moment of sample annealed at 800 °C could not be calculated due to uncertainties in its density and stoichiometry, see section 8 of the Supporting Information for details.

The coercive force $H_C$ of the $(Co,Fe)_{23}B_6$ phase ranges from 70 to 100 Oe. The variations in $H_C$ values could be attributed to several factors, including changes in film thickness (Table S1.1), deformation, or the size of the crystallites[43] (Table S2.4). Nevertheless, $H_C$ for the $(Co,Fe)_{23}B_6$ films is significantly higher than that of the amorphous phase ($H_C \sim 4$ Oe) and lower than that of CoFe ($H_C \sim 250$ Oe).[44]

The loop squareness ($M_R/M_S$) also strongly depends on the crystal structure of the film. $(Co,Fe)_{23}B_6$ exhibit the highest performance ($M_R/M_S \sim 0.83$) compared to amorphous CoFeB or crystallized CoFe ($M_R/M_S \sim 0.6$). Measurements of $M_R$ and $M_S$ as a function of the rotation angle in the plane of the film did not show any anisotropy (details are shown in section 9 of the Supporting Information).

Our results demonstrate that CoFeB films, upon crystallization into the nanosized phase $(Co,Fe)_{23}B_6$, show improved magnetic properties, that are superior to those of amorphous CoFeB, while also presenting benefits over crystalline CoFe. These findings further encourage a deeper exploration of the crystallization mechanisms of CoFeB.

We also investigated the temperature dependencies $M(T)$ for samples annealed at 300 °C (amorphous CoFeB, sample STa-1) and 500 °C (($Co,Fe)_{23}B_6$, sample STa-4) (**Figure 6**).

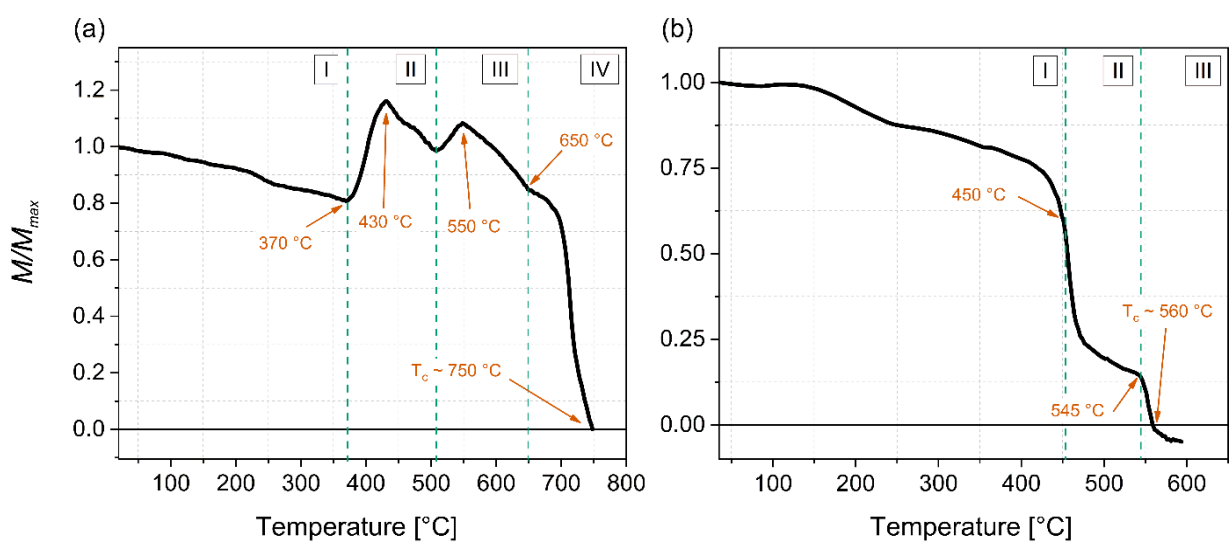

**Figure 6.** Temperature dependence of normalized magnetization $M/M_{MAX}$ for samples (a) STa-1 (annealed at 300°C, amorphous CoFeB phase) and (b) STa-4 (annealed at 500°C, $(Co,Fe)_{23}B_6$ phase).

For sample STa-1, the $M(T)$ curve demonstrates a complex behavior (Figure 6a). In region (I), the magnetization decreases until it reaches a minimum at 370 °C, which is expected due to thermal demagnetization and the potential onset of film oxidation. At the beginning of region (II), a sharp increase in magnetization is observed, reaching a local maximum at 430 °C. We attribute this behavior to the partial crystallization of the amorphous CoFeB film into CoFe, which, as mentioned previously, exhibits higher magnetization.[41,45] Also, CoFe cluster crystallization induces boron out-diffusion from CoFeB.[10,46] Following the local maximum, region (II) shows a decrease in magnetization to the next local minimum at 510 °C, probably due to enhanced Co or Fe oxidation and gradual grain growth.[10] In region (III), we observe another local maximum at 550 °C, which we associate with the crystallization of $(Co,Fe)_2B$ clusters[39] that may locally enhance magnetization. The subsequent decrease in magnetization in region (IV) corresponds to the gradual transition of the entire system to a paramagnetic state. The magnetization reaches zero at 750 °C, which is close to the Curie temperatures ($T_C$) of CoFe and $Fe_2B$, thereby supporting our interpretation.[35,37] A small dip at 650 °C may result from the formation of weakly magnetic silicides ($Fe_3Si$ or $FeSi$) due to excessive film diffusion into the substrate.[47]

The $M(T)$ curve for sample STa-4 demonstrates a simpler behavior (Figure 6b). Similar to the amorphous sample, the initial demagnetization occurs in region (I). However, at 450 °C, a sharp drop in magnetization is observed, followed by a plateau in region (II) until the next abrupt decrease at 545 °C. One possible explanation for the first drop at the (I)-(II) boundary could be a partial demagnetization of the $Fe_{23}B_6$ fraction, suggesting that its $T_C \sim 470$ °C. The subsequent plateau and the second sharp drop in magnetization may be associated with the complete demagnetization of the remaining material (in particular, the $Co_{23}B_6$ fraction), which corresponds to a $T_C$ of 560 °C.[48,49]

The results derived from $M(H)$ and $M(T)$ measurements demonstrate significant differences in the magnetic properties of the exotic $(Co,Fe)_{23}B_6$ phase when compared to amorphous CoFeB or crystallized CoFe in thin films. Importantly, the magnetic measurements show a high level of consistency with the results of structural characterization and align fully with previously published reports.

## 3. Conclusion

This work demonstrates the successful crystallization of amorphous CoFeB thin films, produced through pulsed laser deposition, into the metastable $(Co,Fe)_{23}B_6$ phase. The formation of this phase was confirmed through detailed XRD analysis. The $(Co,Fe)_{23}B_6$ films crystallize with a (111) texture, with the most prominent (333) reflection occurring at $2\theta \approx 45°$ (using Cu Kα radiation). Distinct Laue oscillations in the XRD patterns confirm the exceptional crystal quality of the samples. The $(Co,Fe)_{23}B_6$ phase can form under a wide range of conditions, including various film thicknesses, cooling rates, temperatures and substrate types. On Si/SiO$_2$ substrates, this phase starts to crystallize at 400 °C and remains the dominant one until 750 °C. This confirms that even at elevated temperatures, the diffusion of boron into SiO$_2$ or towards the interface is minimal. At even higher temperatures, we detected the formation of (Co,Fe)$_2$B and CoFe. In contrast, on MgO substrates, (Co,Fe)$_{23}$B$_6$ can only be stabilized at lower temperatures, around 500 °C, while higher temperatures lead to the crystallization of epitaxial CoFe. We suppose that the critical factor for the formation of this exotic phase in our experiments is the stoichiometric transfer of boron from the Co$_{40}$Fe$_{40}$B$_{20}$ target during the PLD process, which facilitates the retention of approximately 20 at.% boron in the film, essential for the formation the boron-rich τ-boride structure.

The (Co,Fe)$_{23}$B$_6$ films demonstrate superior magnetic properties compared to amorphous CoFeB, including a higher saturation magnetization $M_S$ (1400 – 1450 emu cm$^{-3}$), enhanced coercivity $H_C$ (~100 Oe), and improved ratio $M_R/M_S$ ~ 0.83. In certain aspects, this phase can outperform even CoFe. These characteristics, coupled with the compatibility with standard fabrication processes, make (Co,Fe)$_{23}$B$_6$ a promising material for spintronic applications. Additionally, the potential for epitaxial integration with functional oxides suggests opportunities for multiferroic and magnetoelectric devices. Our findings advance the understanding of CoFeB crystallization pathways and highlight (Co,Fe)$_{23}$B$_6$ as a viable, high-performance alternative to conventional phases in magnetic thin-film technologies.

## 4. Experimental Section/Methods

*Thin film deposition.*

CoFeB films were deposited utilizing a pulsed laser deposition (PLD) setup. For this process, we employed a Nd:YAG laser equipped with harmonic splitters to produce radiation at a wavelength of 266 nm. The energy density of the laser pulse, the repetition rate of the pulses, and the duration of each pulse were 1 J cm$^{-2}$, 10 Hz, and 7 ns, respectively. Thin films were deposited from a Co$_{40}$Fe$_{40}$B$_{20}$ target, which had a purity of 99.99%. Annealing of the structures was performed using a laser heating system (Dr. Mergenthaler GmbH & Co. KG, laser processing head LH501, 915 nm, with high-speed fiber coupled 2 color pyrometer), with temperatures $T_a$ ranging from 300 to 800 °C. Both deposition and annealing processes occurred under high vacuum conditions, with a base pressure of 10$^{-6}$ Pa. Additionally, we used two different types of commercial substrates: (100)-oriented single-crystal silicon (CJSC Telecom-STV) with a thermal oxide layer (oxide thickness ~ 250 nm) and monocrystalline (100)-oriented MgO (CrysTec GmbH).

For transmission electron microscopy (TEM) measurements, a special CoFeB sample (S-Mo-TEM) was prepared, which included an additional molybdenum capping layer. The CoFeB was produced as detailed earlier, and the molybdenum was deposited at room temperature after the CoFeB layer underwent annealing and was then cooled in a vacuum for 40 minutes. This capping layer, which was about 8 nm thick, served to protect the CoFeB from oxidation while being transported to the TEM equipment.

Using the same recipe, samples with molybdenum and silicon protective top layers (designated SMag-Mo and SMag-Si) were prepared for additional magnetic measurements.

*Structural investigations.*

The elemental composition of the produced films was analyzed through Rutherford backscattering spectrometry (RBS). This analysis was conducted at the Van de Graaff accelerator AN2500 (High Voltage Engineering Europa B.V.) with an incident $^4He^+$ beam energy of 1.4 MeV. The angle between the incident beam and the sample's surface normal was 2°, while the scattering and exit angles were 170° and 10°, respectively. The experimental RBS spectra were fitted with using Simnra software (version 7.03).[50] Additionally, the samples were analyzed using energy dispersive X-ray spectroscopy (EDX) employing an Oxford Instruments X-ACT analyzer integrated within a Jeol JSM-6390 scanning electron microscope. The electron beam energy was set to 20 keV, and the spectra were processed using Inca Energy software (version 5.05).

X-ray diffraction patterns of the produced films were obtained using a Bruker AXS D8 DISCOVER system with Cu Kα (0.15406 nm) radiation. Initially, the samples were analyzed in Bragg-Brentano (θ–2θ) geometry. However, for highly-textured films, this method alone is insufficient for accurate phase identification. Consequently, further measurements were conducted with a tilted sample (as we previously did for various complex materials[51,52]) or in an asymmetrical configuration[53] (refer to Supporting Information, section 10, for further details). X-ray reflectometry measurements were also performed using a Bruker D8 diffractometer in conventional θ–2θ geometry. The reflectometry curves were analyzed using the open-source software GenX 3.[54]

The surface morphology was investigated using an atomic force microscope (AFM) SmartSPM Aist-NT. Commercial cantilevers (TipsNano Co) NSG10 and NSG30 with a console length of 125 μm were used for operation in the dynamic mode, with an amplitude in the range of 2-10 nm.

A cross-sectional TEM lamella was prepared from the Mo coated specimen using standard FIB lift-out technique on a Thermo Fischer Helios NanoLab 650 dual beam system.

Final polishing of the lamella was performed at 2 keV Ga energy to minimize ion induced damage. TEM imaging was performed using a JEOL ARM200F TEM operating at 200 kV.

*Magnetic measurements.*

The field dependence of magnetization $M(H)$ was measured using a vibrating sample magnetometer (Lakeshore 7400 System) at room temperature in the field up to 3 kOe. The temperature dependence of magnetization $M(T)$ was obtained using a high-temperature oven (model 74034) up to 800 °C at a rate of 5 °C min$^{-1}$. The measurements were carried out in a magnetic field of 200 Oe, and the sample was blown with argon at a flow rate of 100 sccm.

**Funding**

This work was prepared with support from the Ministry of Science and Higher Education of the Russian Federation (project FZWM-2024-0011).

**Acknowledgements**

The authors would like to express their gratitude to Alexey Grunin for his valuable contribution to the discussion and assistance in interpreting the results obtained.

**Data Availability Statement**

The datasets generated and/or analyzed during the current study are available in the Figshare repository, https://doi.org/10.6084/m9.figshare.30657671

Supporting Information

# Structure and Magnetic Properties of Vacuum-Annealed CoFeB Thin Films: From Amorphous Alloy to Metastable (Co,Fe)$_{23}$B$_6$ τ-Boride

*Petr Shvets\*, Grigorii Kirichuk, Vitalii Salnikov, Jacques O'Connell, Valeria Rodionova, Alexandr Goikhman, Ksenia Maksimova*

## 1. List of studied samples

**Table S1.1.** List of samples produced for the current study. $T_a$ is the annealing temperature, $CR$ is the cooling rate. The thickness was determined using X-ray reflectivity ($h_{XRR}$, refer to **Figure S1.1**) and Laue oscillations ($h_{Laue}$).

| Sample code | Substrate | $T_a$ [°C] | $CR$ [°C/min$^{-1}$] | $h_{XRR}$ [nm] | $h_{Laue}$ [nm] |
|---|---|---|---|---|---|
| STh-1 | Si / SiO$_2$ | 500 | 200 | 5.5 | 5.9 |
| STh-2 | | | | 9.5 | 9.7 |
| STh-3 | | | | 21.5 | 22.1 |
| STh-4 | | | | 26.0 | 26.7 |
| S-Mo-TEM | | | | 12.3 | 12.1 |
| SMag-Mo | | | | 12.6 | 11.9 |
| SMag-Si | | | | 12.5 | 11.9 |
| STa-1 | Si / SiO$_2$ | 300 | 200 | 27.5 | – |
| STa-2 | | 350 | | 11.2 | – |
| STa-3 | | 400 | | 20.0 | 20.3 |
| STa-4 | | 500 | | 21.5 | 22.1 |
| STa-5 | | 600 | | 15.8 | 16.0 |
| STa-6 | | 700 | | 17.5 | 17.4 |
| STa-7 | | 750 | | 13.0 | 13.1 |
| STa-8 | | 800 | | 14.5 | – |
| SCr-1 | Si / SiO$_2$ | 500 | 200 | 9.5 | 9.7 |
| SCr-2 | | | 50 | 11.8 | 11.8 |
| SCr-3 | | | 10 | 8.5 | 8.5 |
| SCr-4 | | | 2 | 10.3 | 10.5 |
| SMg-1 | MgO | 500 | 200 | – | 11.0 |
| SMg-2 | | 600 | | – | 11.0 |
| SMg-3 | | 700 | | – | 11.0 |

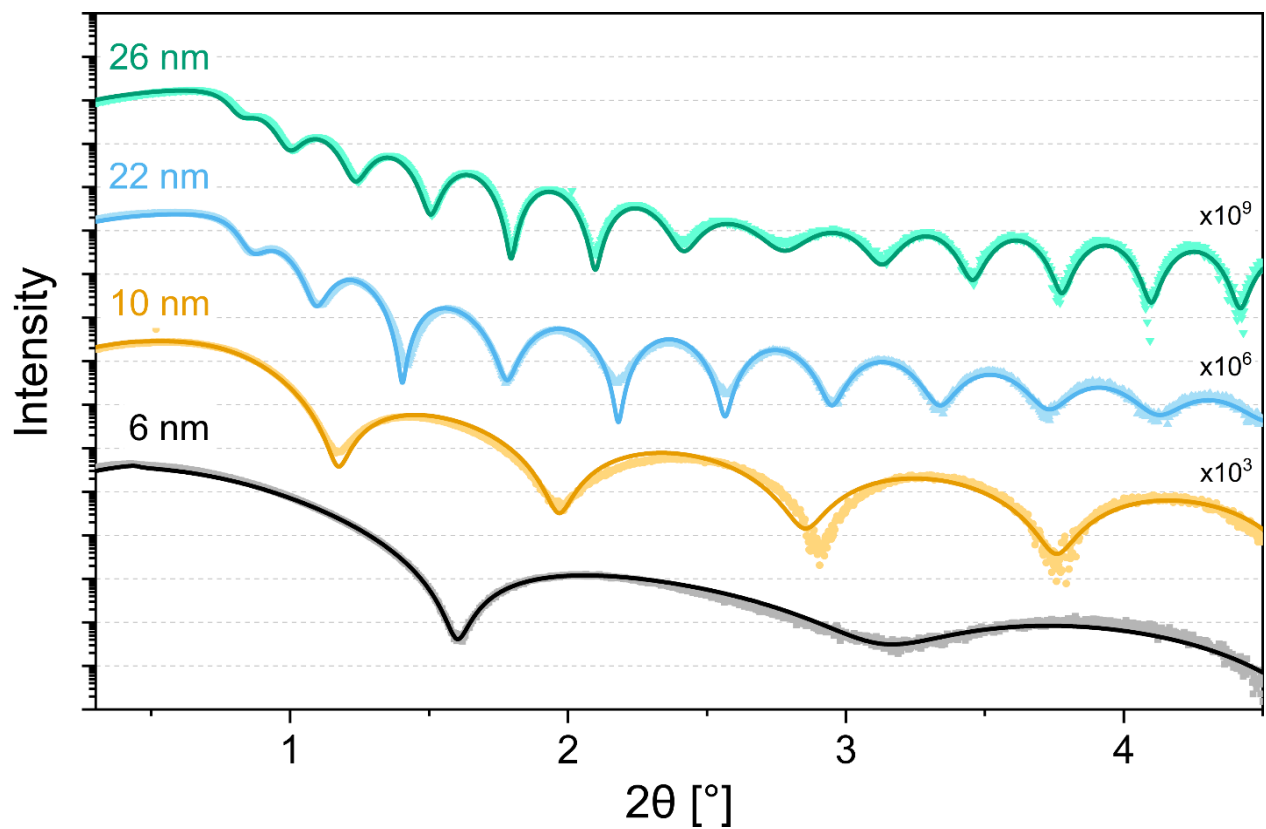

**Figure S1.1.** XRR curves for CoFeB films. The solid curves represent the optimal fits obtained with the specified thicknesses. The graphs are presented in a logarithmic scale and offset for clarity.

## 2. XRD of CoFeB films

**Table S2.1.** Conditions required for the observation of (hkl) peaks in the $(Co,Fe)_{23}B_6$ crystal with (111) texture (angles $\omega$ and $\psi$) and positions of the detected peaks ($2\theta$) for sample STh-4.

| (hkl) | $\omega$ [°] | $\psi$ [°] | $2\theta$ [°] |
|---|---|---|---|
| (111) | 2θ/2 | 0 | 14.64 |
| (222) | 2θ/2 | 0 | 29.45 |
| (333) | 2θ/2 | 0 | 44.81 |
| (444) | 2θ/2 | 0 | 61.09 |
| (555) | 2θ/2 | 0 | 78.83 |
| (666) | 2θ/2 | 0 | 99.25 |
| (400) | 2θ/2 | 54.7 | 34.0 |
| (420) | 2θ/2 | 39.2 | 38.2 |
| (422) | 1.6 | 0 | 42.11 |
| (511) | 2θ/2 | 39 | 44.69 |
| (440) | 2θ/2 | 35.3 | 48.89 |
| (531) | 2θ/2 | 28.6 | 51.34 |
| (600) | 2θ/2 | 54.7 | 51.99 |
| (442) | 10.3 | 0 | 52.21 |
| (533) | 14.3 | 0 | 57.46 |
| (622) | 2θ/2 | 29.5 | 58.2 |
| (800) | 2θ/2 | 54.7 | 71.56 |
| (644) | 25.8 | 0 | 74.35 |
| (660) | 3.1 | 0 | 76.81 |
| (751) | 9.4 | 0 | 78.73 |
| (840) | 1.6 | 0 | 81.72 |
| (753) | 23.8 | 0 | 83.79 |
| (931) | 6.2 | 0 | 88.55 |
| (844) | 26.4 | 0 | 91.76 |
| (933) | 17.3 | 0 | 93.55 |
| (953) | 28.0 | 0 | 103.54 |
| (10 4 0) | 10.6 | 0 | 104.06 |
| (11 1 1) | 6.8 | 0 | 108.35 |

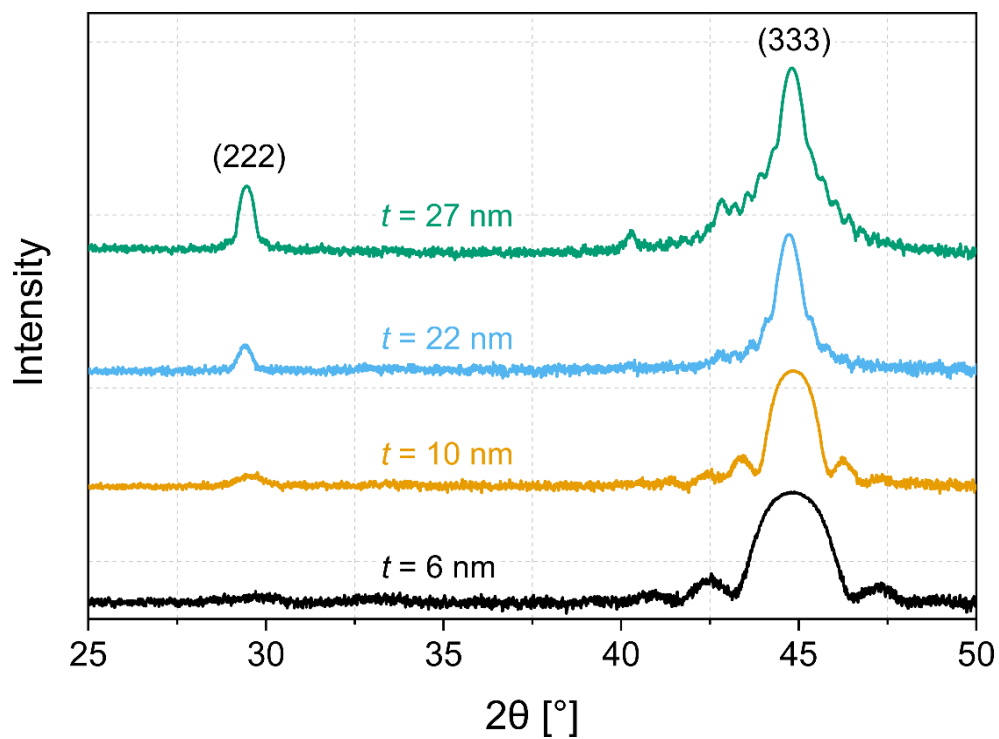

**Figure S2.1.** XRD patterns for annealed CoFeB films with different thicknesses (samples STh1 – STh4). The graphs are presented in a logarithmic scale and offset for clarity.

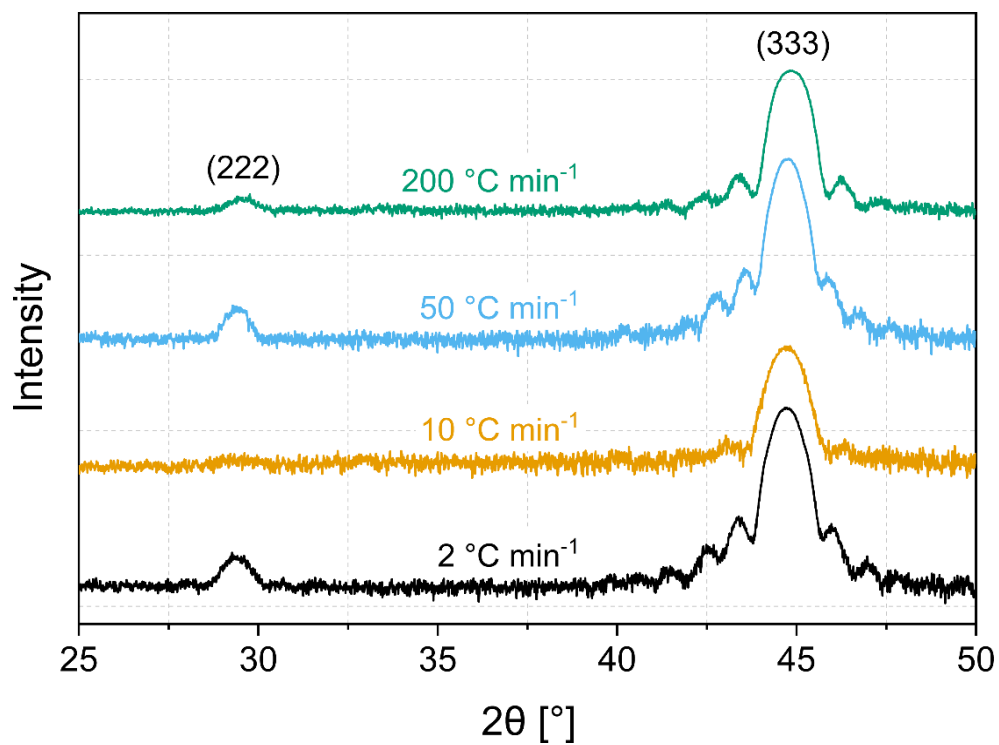

**Figure S2.2.** XRD patterns for annealed CoFeB films with different cooling rates (samples SCr1 – SCr4). The graphs are presented in a logarithmic scale and offset for clarity.

**Table S2.2.** Conditions required for the observation of (hkl) peaks in the $(Co,Fe)_2B$ crystal with (002) texture (angles ω and ψ) and positions of the detected peaks (2θ) for sample STh-4.

| (hkl) | ω [°] | ψ [°] | 2θ [°] |
|---|---|---|---|
| (002) | 2θ/2 | 0 | 42.93 |
| (004) | 2θ/2 | 0 | 94.01 |
| (211) | 2θ/2 | 61.7 | 45.08 |
| (202) | 2θ/2 | 39.5 | 56.74 |
| (213) | 8.0 | 0 | 80.29 |
| (413) | 7.5 | 0 | 112.45 |

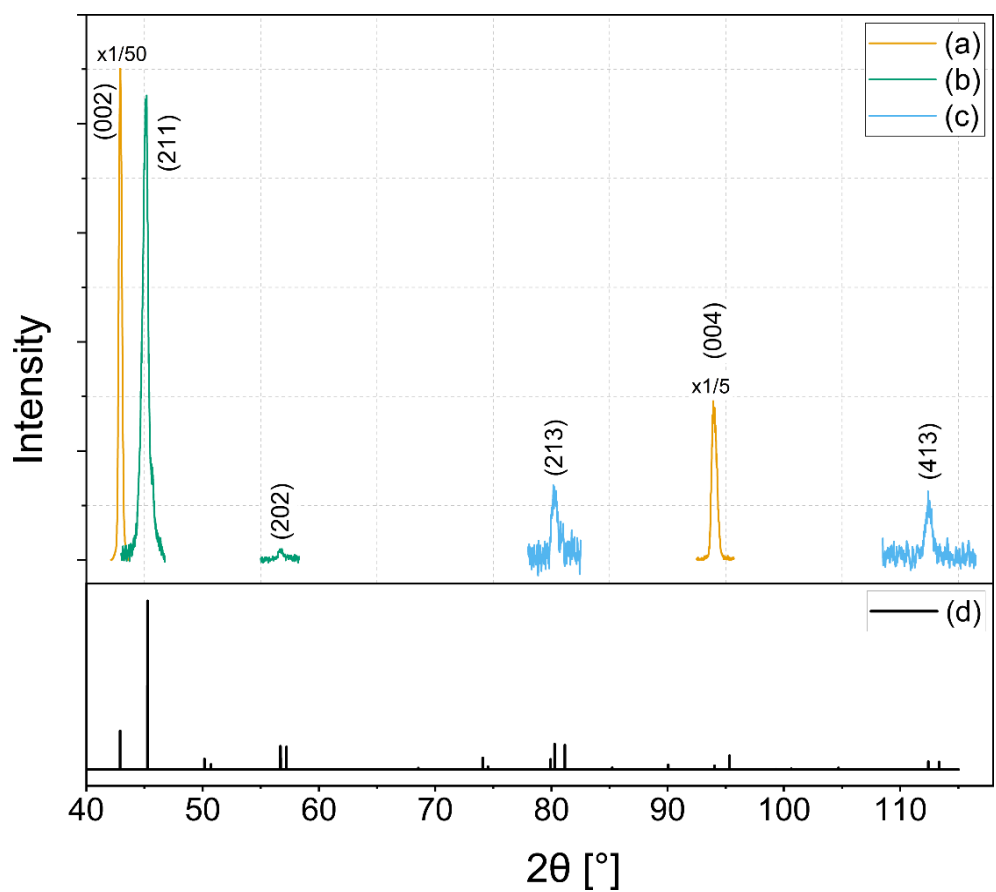

**Figure S2.3.** XRD patterns for sample STa-8: (a) θ–2θ geometry, (b) tilted sample geometry, (c) asymmetrical geometry. (d) Theoretical XRD pattern of $M_2B$ powder computed using positional parameters from literature,[1] and lattice constants of $a = 5.088$ Å, $c = 4.212$ Å.

**Table S2.3.** Conditions required for the observation of (hkl) peaks in the (Co,Fe) crystal with (110) and (200) textures (ω and ψ) and positions of the detected peaks (2θ) for sample STa-8.

| (hkl) | ω [°] | ψ [°] | 2θ [°] |
|---|---|---|---|
| (110) texture | | | |
| (110) | 2θ/2 | 0 | 45.01 |
| (200) | 2θ/2 | 45 | 65.34 |
| (221) | 11.5 | 0 | 82.88 |
| (220) | 2θ/2 | 0 | 99.77 |
| (200) texture | | | |
| (110) | 2θ/2 | 45 | 45.07 |
| (200) | 2θ/2 | 0 | 65.62 |
| (221) | 6.2 | 0 | 83.06 |

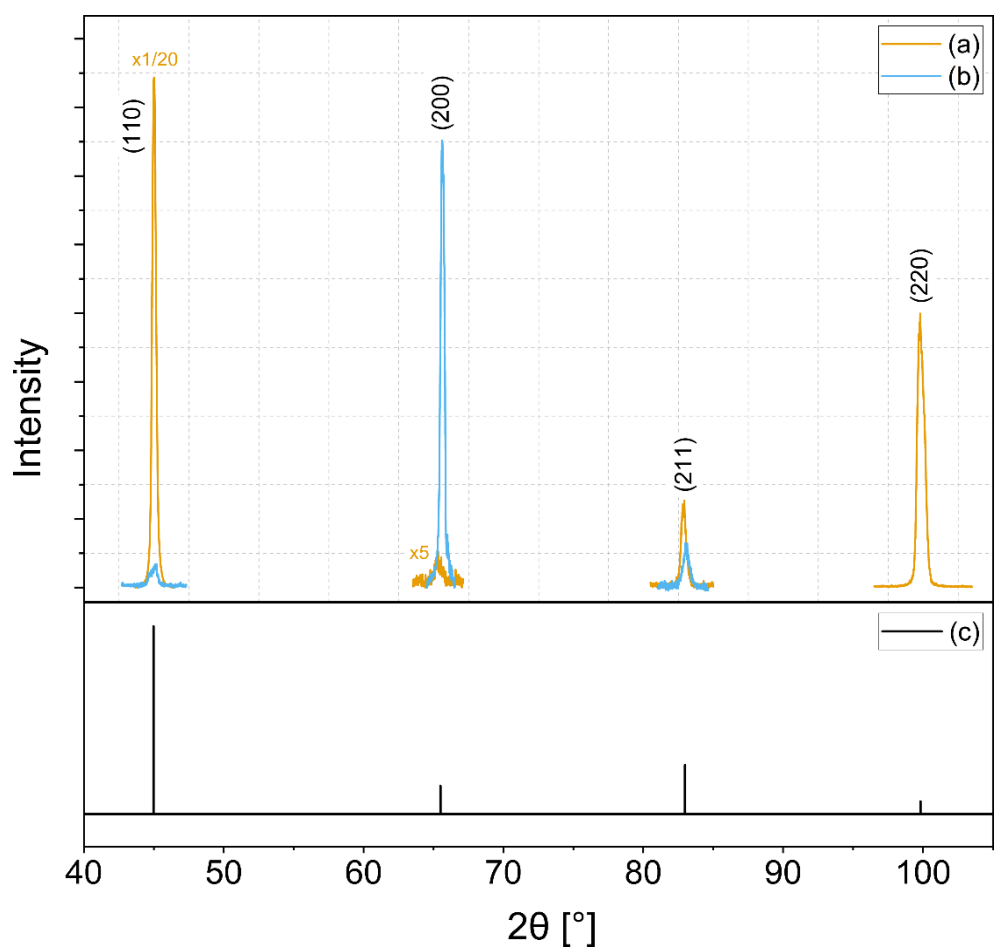

**Figure S2.4.** XRD patterns for the CoFe phase of sample STa-8: (a) (110) texture, (b) (200) texture. (c) Theoretical XRD pattern for the CoFe powder with a simple body-centered structure calculated with a lattice constant of $a = 2.848$ Å.

**Table S2.4.** The crystallite size (*CS*) and strain (ε) calculated using the Williamson-Hall method, and thickness ($h_{XRR}$) determined by XRR for $(Co,Fe)_{23}B_6$ samples produced at different temperatures.

| Sample code | $T_a$ [°C] | $h_{XRR}$ [nm] | CS [nm] | ε [%] |
|---|---|---|---|---|
| STa-3 | 400 | 20.0 | 20 | 0.02 |
| STa-4 | 500 | 21.5 | 23 | 0.04 |
| STa-5 | 600 | 15.8 | 16 | –0.01 |
| STa-6 | 700 | 17.5 | 17 | –0.02 |
| STa-7 | 750 | 13.0 | 13 | –0.04 |

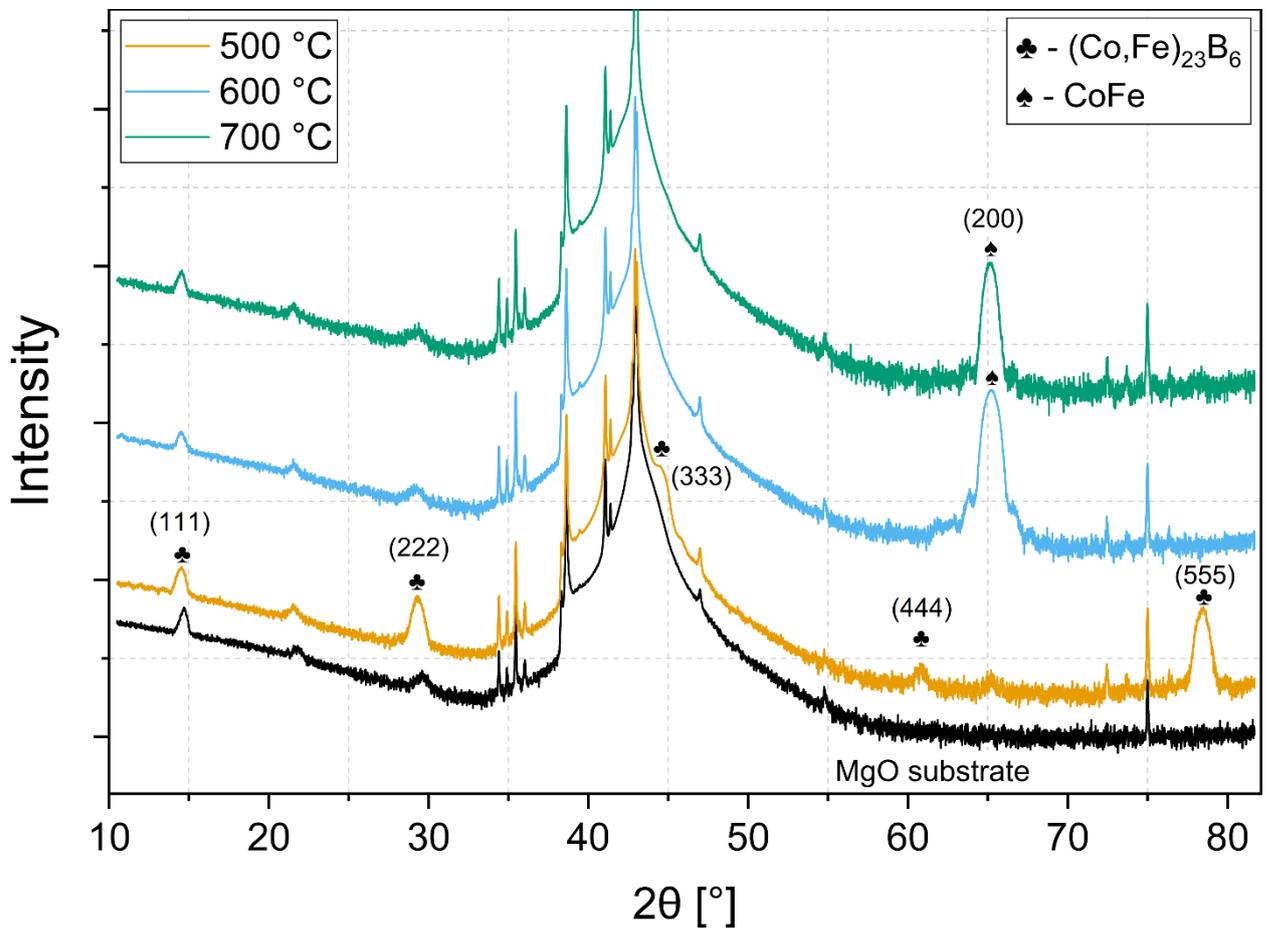

**Figure S2.5.** XRD patterns for FeCoB films deposited on an MgO substrate annealed at different temperatures (samples SMg-1 – SMg-3). The graphs are presented in a logarithmic scale and offset for clarity. Unmarked peaks belong to the MgO substrate.

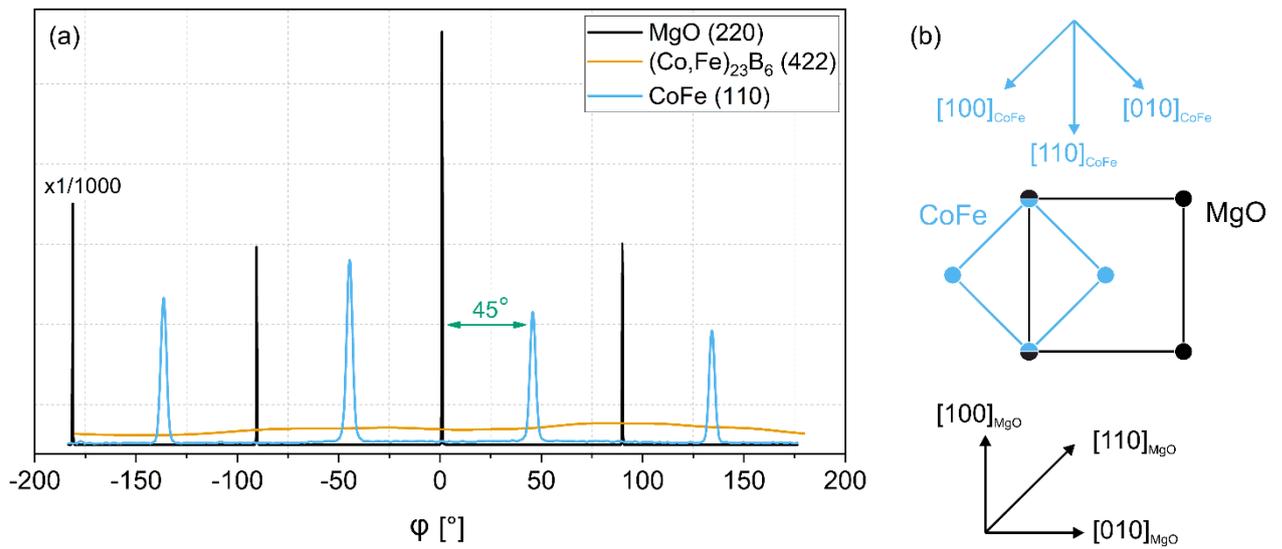

**Figure S2.6.** (a) φ-scans for the (110) CoFe reflection (ω = 22.5°, 2θ = 45°, ψ = 45°), the (422) (Co,Fe)$_{23}$B$_6$ reflection (ω = 1.8°, 2θ = 42.1°, ψ = 0°), and the (220) substrate reflection (ω = 31.15°, 2θ = 62.3°, ψ = 45°). (b) Diagram illustrating the mutual arrangement of the crystal lattices of MgO and CoFe for epitaxial films.

## 3. Surface morphology for (Co,Fe)$_{23}$B$_6$ samples produced at different temperatures

The surface of samples STa-1, STa-3 – STa-6, and STa-8 was investigated using atomic force microscopy (AFM) (**Figure S3.1**). At an annealing temperature of 300 °C, the film has an amorphous structure along with a smooth and uniform surface morphology. As the annealing temperature increases, grains start to appear on the surface (400 °C), which subsequently develop into distinct small clusters (500 °C), and eventually lead to the formation of separate large grains with a negligible number of combined clusters (600 °C). At 700 °C, the grains converge into a textured uniform film. According to the X-ray diffraction data (Figure 2 of the main text), the films are highly-textured at annealing temperatures between 400 and 700 °C. This suggests that the nucleation and growth of grains may occur in a preferentially oriented manner. At an annealing temperature of 800 °C, a partial degradation of the film is observed, where only large clusters, presumably of CoFe, remain on the surface.

It is important to note that the CoFeB metal films are highly sensitive to oxidation in the atmosphere, which can have a considerable impact on the surface morphology, leading to a potential misrepresentation of the actual behavior of the thin film.

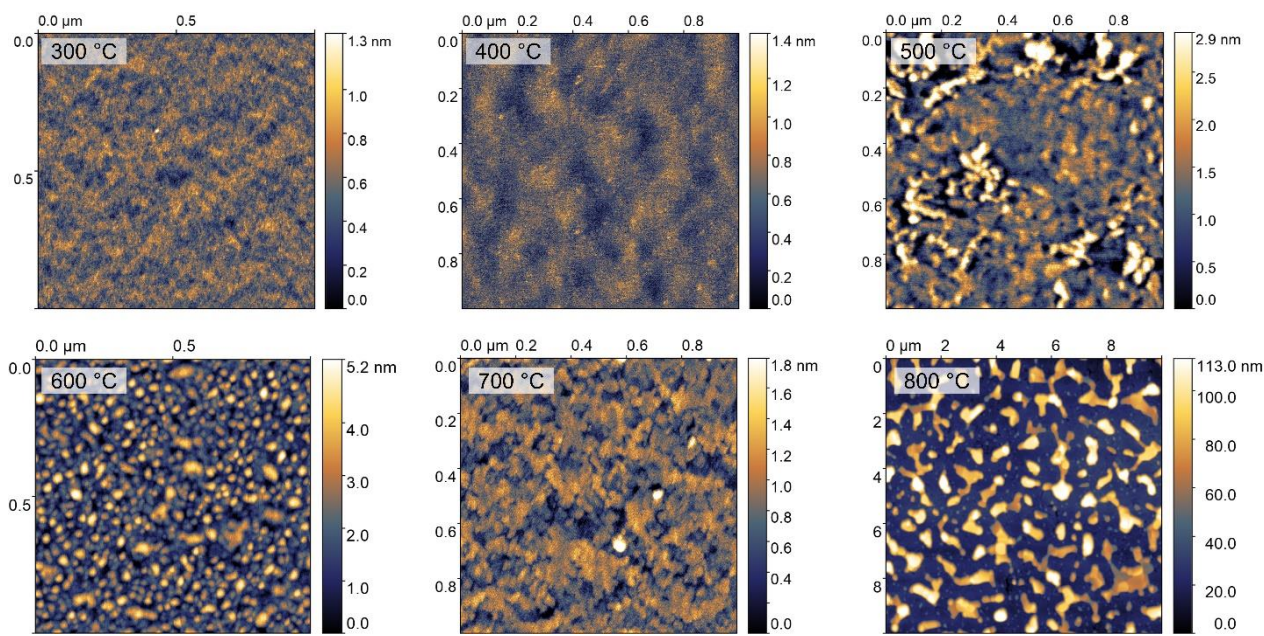

**Figure S3.1.** AFM scans for (Co,Fe)$_{23}$B$_6$ films produced at different temperatures.

## 4. Study of the magnetic properties of CoFeB thin films grown on MgO substrates

The hysteresis loops for the samples produced on MgO are shown in **Figure S4.1**, with the magnetic properties summarized in **Table S4.1**. The obtained values are consistent with the findings from the structural studies detailed in Section 2 of the main text. When CoFe is formed in the film (annealing temperatures of 600 and 700 °C), the saturation magnetization increases to 1700-1800 emu cm$^{-3}$, while for the $(Co,Fe)_{23}B_6$ phase, it is about 1400 emu cm$^{-3}$, which aligns with the results observed for samples on silicon substrates. A further interesting observation is that for the sample annealed at 700 °C, the coercivity and the "squareness" of the loop ($M_R/M_S$) achieve their maximum value, which may be attributed to the high quality of the epitaxial structure of CoFe.

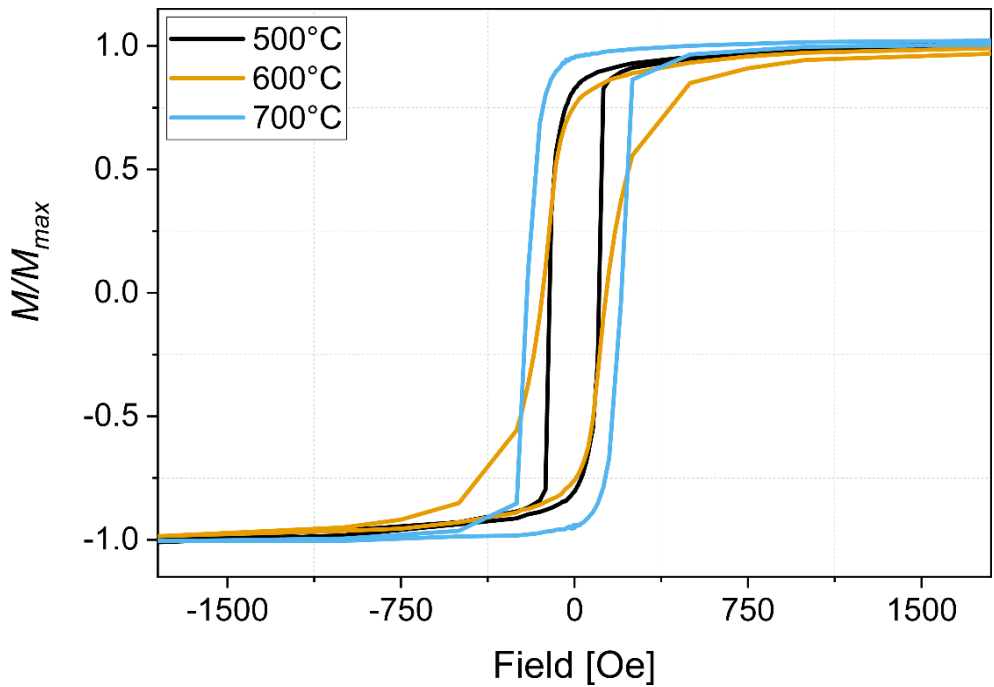

**Figure S4.1.** Hysteresis loops for thin CoFeB films grown on MgO substrates annealed at different temperatures.

**Table S4.1.** The values of saturation magnetization ($M_S$), coercivity ($H_c$) and loop squareness ($M_R/M_S$) for the samples SMg-1 – SMg-3.

| Sample code | $T_a$ [°C] | $M_S$ [emu cm$^{-3}$] | $H_c$ [Oe] | $M_R/M_S$ |
|---|---|---|---|---|
| SMg-1 | 500 | 1470 | 11 | 0.80 |
| SMg-2 | 600 | 1730 | 140 | 0.76 |
| SMg-3 | 700 | 1820 | 200 | 0.94 |

## 5. Magnetocrystalline anisotropy of CoFe(B) on MgO

For thin CoFeB films grown on MgO substrates, the in-plane magnetic anisotropy was also investigated. The film in which the $(Co,Fe)_{23}B_6$ phase was formed exhibited no anisotropy, consistent with its uniform φ-scan (Figure S2.6) (results are not shown). However, when an epitaxial CoFe film was formed on MgO, an in-plane anisotropy appeared (**Figure S5.1**). During the measurements, the [100] axis of the MgO substrate was initially aligned with the magnetic field vector, corresponding to a rotation angle of 0°. It is important to note that the easy axis of magnetization is oriented along the edge of the substrate: [100] or [010] direction. In section 2, we demonstrated that the CoFe film was formed on a substrate with a rotation of 45°. Consequently, the [100] direction of MgO corresponds to the [110] direction in CoFe (Figure S2.6). Therefore, the edge of CoFe is the hard axis of magnetization, while the diagonal is the easy axis. This behavior is uncommon for such structures, although there are reports on the formation of such anisotropy in epitaxial Fe films on MgO substrates.[2] In our work, we attribute this behavior to the indirect formation of CoFe, which results from the out-diffusion of boron from the CoFeB film.

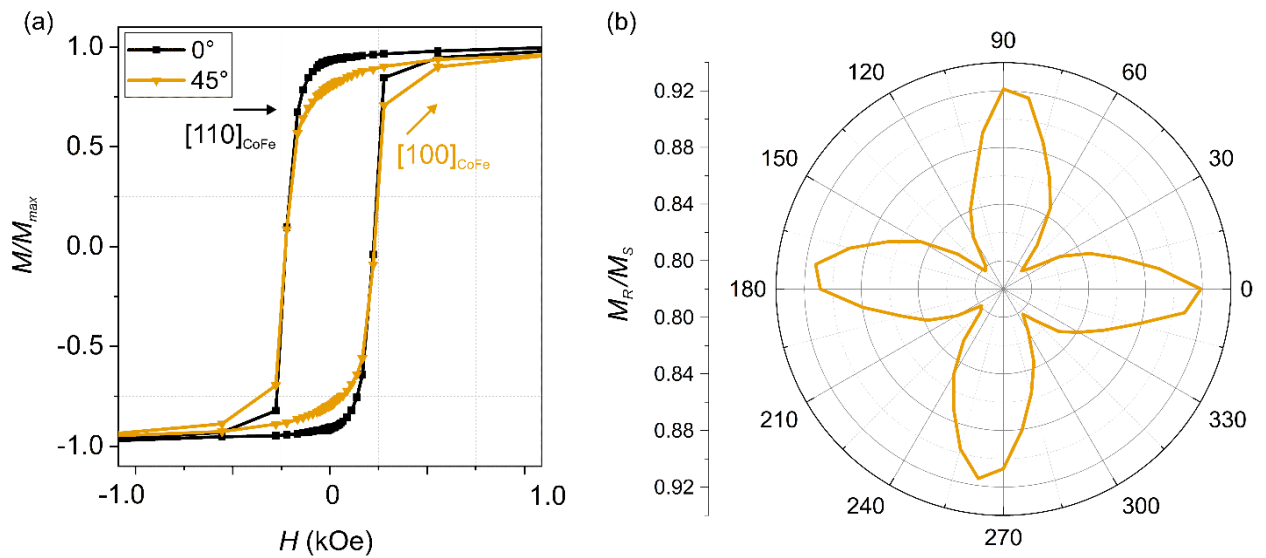

**Figure S5.1.** (a) Hysteresis loops for the epitaxial CoFe film measured at a rotation angle of 0°: [100] for MgO and [110] for CoFe; and at 45°: [110] for MgO and [100] for CoFe. (b) The dependance of squareness $M_R/M_S$ on the azimuthal angle showing the uniaxial anisotropy behavior.

## 6. Influence of capping layers on magnetic characteristics

Here we study the influence of a capping layer on the magnetic properties of $(Co,Fe)_{23}B_6$ films. Silicon and molybdenum were chosen as the capping layers due to their ability to protect the CoFeB metal film from surface oxidation. The hysteresis loops (**Figure S6.1**) for samples with different capping layers are almost identical. A comparison of coated and uncoated $(Co,Fe)_{23}B_6$ samples reveals that the loop shape, saturation magnetization, and coercivity do not exhibit significant changes. Therefore, it can be concluded that the measurements of samples with a surface oxide layer (typically 5-10 Å according to XRR data) are relevant and accurate.

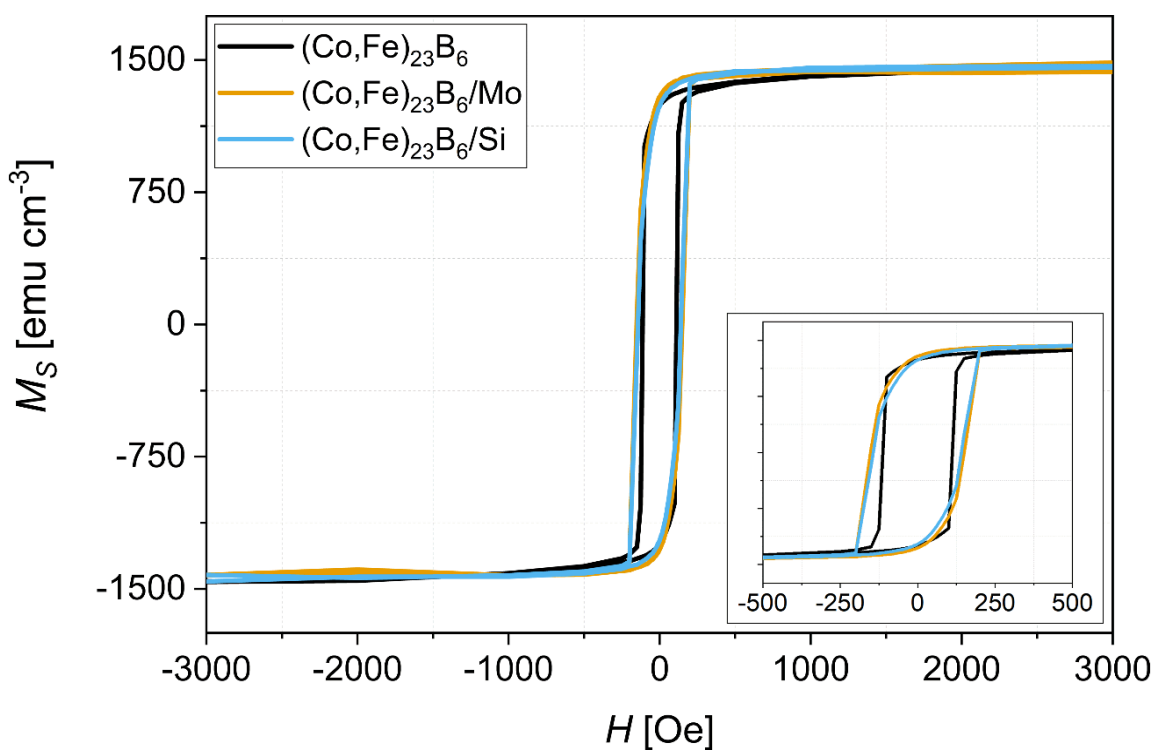

**Figure S6.1.** Hysteresis loops for $(Co,Fe)_{23}B_6$ samples with Mo and Si capping layers, as well as for a sample without a top layer.

## 7. Calculation of the average magnetic moment for Co/Fe atoms

This section describes the method for calculating the average magnetic moment expressed in [$\mu_B$/atom]. First, the magnetic moment is computed in units of Bohr magneton per unit formula, $\mu_{f.u.}(\mu_B)$.[3]

Magnetization $M$ is defined as the magnetic moment ($m_{total}$) per volume ($V$):

$$M = \frac{m_{total}}{V} \tag{7.1}$$

The total magnetic moment can be expressed as:

$$m_{total} = N_{f.u.} \cdot \mu_{f.u.}, \tag{7.2}$$

where $N_{f.u.}$ is the number of formula units in the volume $V$, $\mu_{f.u.}$ is the magnetic moment per formula unit (in [emu], 1 $\mu_B = 9.274 \cdot 10^{-21}$ emu).

The number of formula units in the volume $V$ is:

$$N_{f.u.} = \frac{m}{M_{molar}} \cdot N_A = \frac{\rho V}{M_{molar}} \cdot N_A, \tag{7.3}$$

where $m$ is the mass of the substance within the volume $V$, $M_{molar}$ is the molar mass of the formula unit [g mol$^{-1}$], $N_A$ is Avogadro's number, and $\rho$ is the density [g cm$^{-3}$]. Combining Equation (7.1), (7.2), and (7.3) we obtain:

$$\mu_{f.u.}(\mu_B) = \frac{M \cdot M_{molar}}{\rho \cdot N_A} \cdot 9.274 \cdot 10^{-21} = 1.79 \cdot 10^{-4} \cdot \frac{M \cdot M_{molar}}{\rho}$$

Now, we can calculate the magnetic moment in [$\mu_B$ atom$^{-1}$]. Assuming that the stoichiometry of the film is (Co, Fe)$_{80}$B$_{20}$, we obtain:

$$\mu_{atom}(\mu_B) = \frac{\mu_{f.u.}}{n_{magnetic\ atoms\ in\ f.u.}} = \frac{\mu_{f.u.}}{0.8} = 1.25 \mu_{f.u.}$$

## 8. STa-8 surface study

The surface of sample STa-8 was additionally investigated by a scanning electron microscope (SEM) Jeol JSM-6390, with an accelerating voltage of 30 kV. The film exhibits a non-uniform, island-like structure, consistent with data acquired through AFM (section 3).

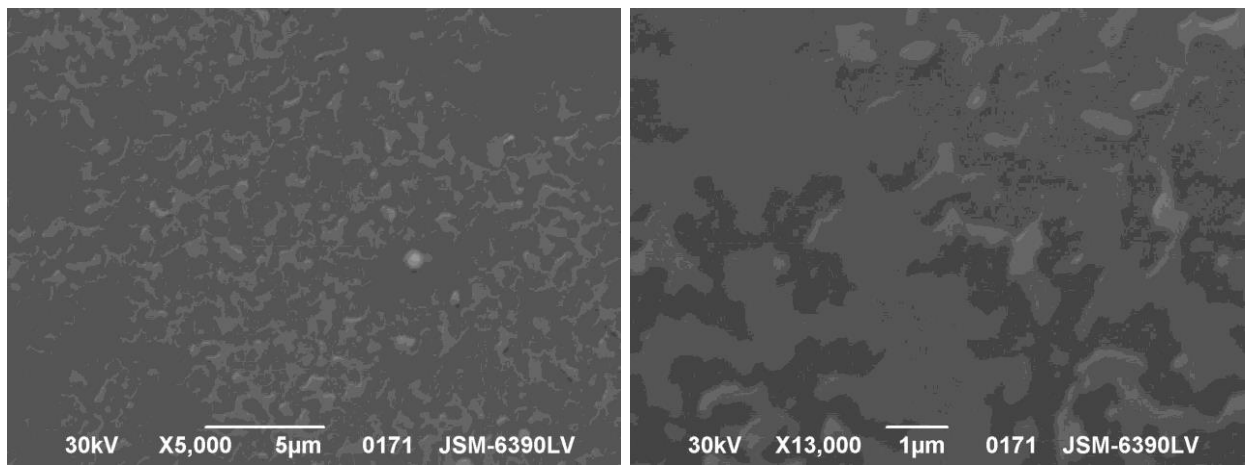

**Figure S8.1.** SEM images of STa-8 surface.

## 9. Study of the magnetic anisotropy

For samples STa-1, STa-3 – STa-6, and STa-8, the magnetic anisotropy "in plane" was investigated. The sample was rotated around its axis within the plane of the magnetic field, and hysteresis loops were recorded with a 30° step. The results reveal a complete absence of anisotropy (**Figure S9.1**). This is in agreement with the structural measurements, as the films, despite being textured, remain polycrystalline, leading to the absence of a defined easy axis of magnetization.

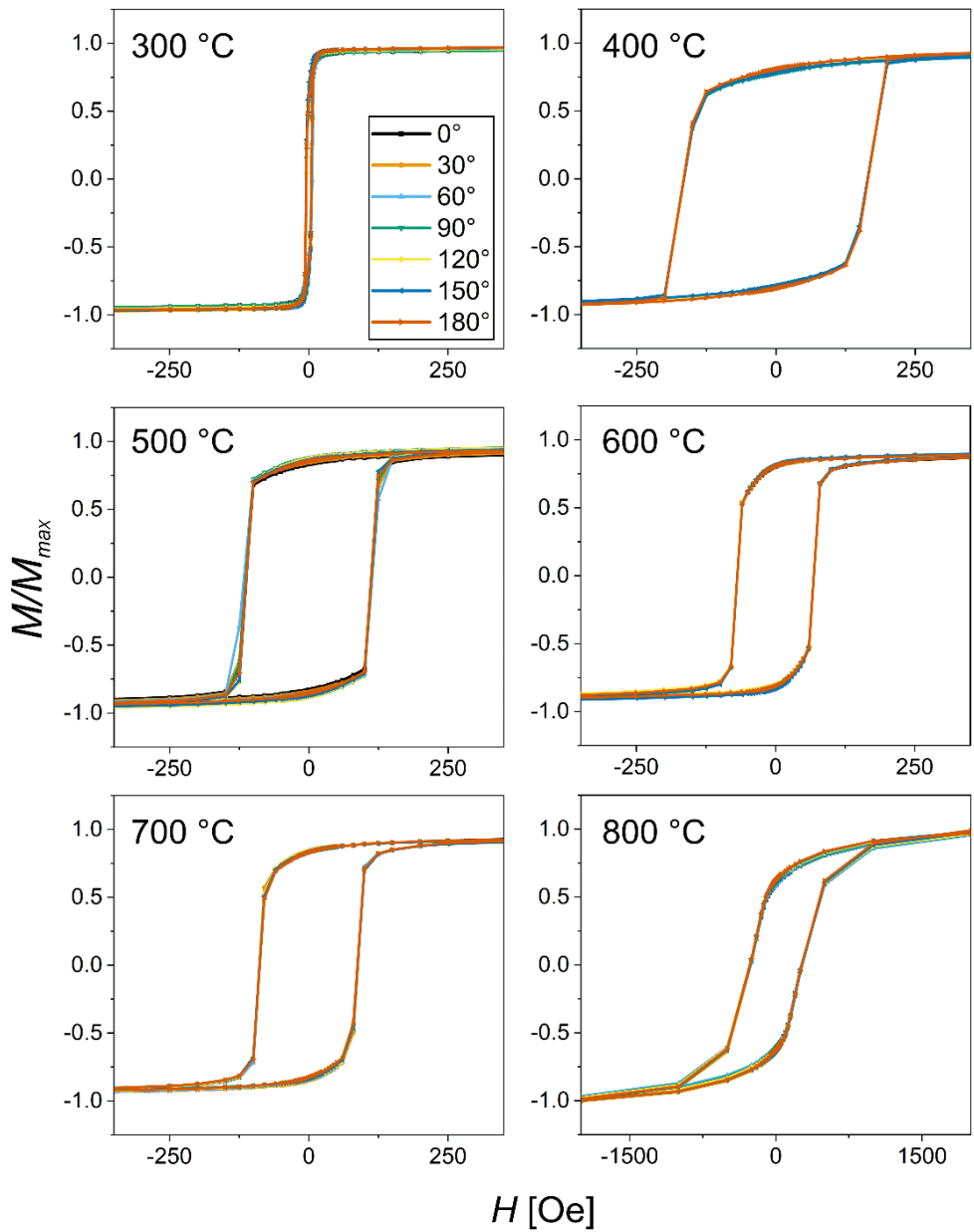

**Figure S9.1.** The dependence of hysteresis loops on the in-plane rotation angle for thin CoFeB films annealed at different temperatures.

## 10. X-ray diffraction investigation of the textured thin films

Let us consider a highly-textured thin film, which has a single peak in the Bragg-Brentano (θ – 2θ) scan. This peak is tentatively assigned to the (*hkl*) reflection of a crystal with lattice constants *a*, *b*, *c*, and the angles *α*, *β*, and *γ*. In order to detect an additional XRD peak corresponding to the crystal, (*h'k'l'*), it is necessary to first compute δ, the angle between the planes (*hkl*) and (*h'k'l'*), utilizing the following equations:[4]

$$\cos\delta = d_{hkl}d_{h'k'l'}[hh'a^{*2} + kk'b^{*2} + ll'c^{*2}$$
$$+ (kl' + lk')b^*c^* \cos\alpha^* + (hl' + lh')a^*c^* \cos\beta^* + (hk' + kh')a^*b^* \cos\gamma^*]$$

where $d_{hkl}$ is an interplanar distance:

$$d_{hkl} = \frac{1}{|h\boldsymbol{a}^* + k\boldsymbol{b}^* + l\boldsymbol{c}^*|},$$

$$|h\boldsymbol{a}^* + k\boldsymbol{b}^* + l\boldsymbol{c}^*|$$
$$= h^2 a^{*2} + k^2 b^{*2} + l^2 c^{*2} + 2klb^*c^* \cos\alpha^* + 2lhc^*a^* \cos\beta^* + 2hka^*b^* \cos\gamma^*,$$

and $a^*, b^*, c^*, \alpha^*, \beta^*,$ and $\gamma^*$ are the lattice constants and angles of the reciprocal lattice:

$$\cos\alpha^* = \frac{\cos\beta \cos\gamma - \cos\alpha}{\sin\beta \sin\gamma},$$

$$\cos\beta^* = \frac{\cos\alpha \cos\gamma - \cos\beta}{\sin\alpha \sin\gamma},$$

$$\cos\gamma^* = \frac{\cos\beta \cos\alpha - \cos\gamma}{\sin\beta \sin\alpha},$$

$$a^* = \frac{bc \sin\alpha}{V}, \quad b^* = \frac{ac \sin\beta}{V}, \quad c^* = \frac{ab \sin\gamma}{V}$$

where $V = abc\sqrt{1 - \cos^2\alpha - \cos^2\beta - \cos^2\gamma + 2\cos\alpha \cos\beta \cos\gamma}$.

Once the value of δ is determined, the sample in the diffractometer should be tilted by this angle (set ψ = δ, refer to **Figure S10.1**) and a standard θ – 2θ scan should be conducted near 2θ that corresponds to the (*h'k'l'*) reflection ($2\theta_{h'k'l'}$). It is important to note that for epitaxial films, this reflection will only be observable for specific in-plane sample orientations (φ-axis). In contrast, for polycrystalline textured films, the in-plane orientations are random, thus the (*h'k'l'*) reflection will be detectable for any φ. However, its intensity will be considerably reduced when compared to the expected intensity of the (*h'k'l'*) peak from a powder sample, which generates the same (*hkl*) signal as the textured film.

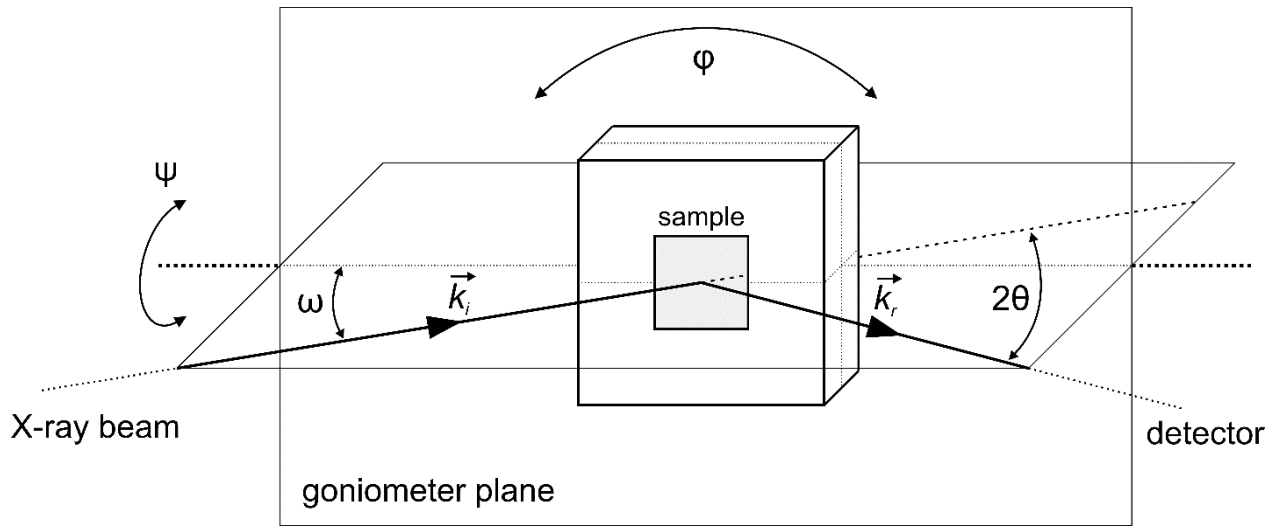

**Figure S10.1.** Geometry of the X-ray diffraction experiment.

An alternative method for probing $(h'k'l')$ planes involves utilizing the asymmetrical scan.[5] In this case, the sample is not tilted ($\psi = 0$). Rather, we fix $\omega = 2\theta_{h'k'l'}/2 - \delta$ and conduct a $2\theta$ scan close to $2\theta_{h'k'l'}$. This setup is advantageous in terms of diffraction intensity (see **Figure S10.2** for a comparison of the two different modes), but it can only be implemented if $2\theta_{h'k'l'}/2 - \delta > 0$.

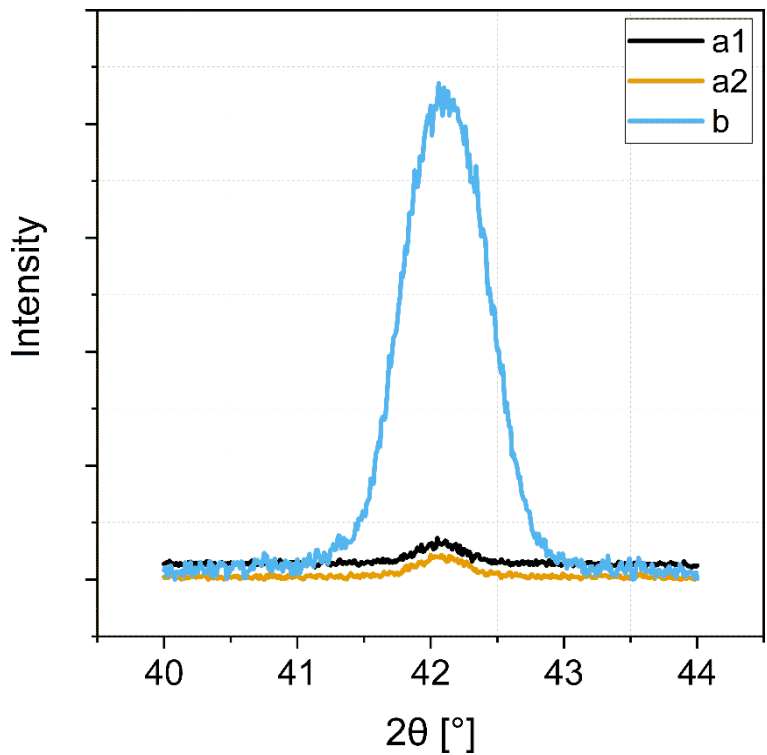

**Figure S10.2.** (422) reflection of the $(CoFe)_{23}B_6$ film with (111) texture measured at (a1) $\psi = 19.5°$, coupled $\theta - 2\theta$ scan, (a2) $\psi = 19.5°$, $\omega = 21.1°$, $2\theta$ scan and (b) $\psi = 0°$, $\omega = 1.6°$, $2\theta$ scan. The notable rise in intensity in asymmetrical scans can be attributed to the following factors:

1) ω is less than 2θ/2 and may approach zero, thereby obtaining all advantages of grazing incidence geometry (effective increase of film thickness).

2) Asymmetrical geometry exhibits a greater tolerance for the alignment of the crystal along the φ-axis. **Figure S10.3** provides an example, where we examined the (311) plane of the MgO (100) monocrystal. A wider φ-scan distribution indicates that in polycrystalline samples, a larger number of randomly oriented crystallites can contribute to the X-ray intensity of the observed reflection.

3) Because of the limitations associated with X-ray optics, the samples that are tilted result in diffraction peaks that have a lower intensity. For instance, at ψ ~ 25°, we lose roughly half of the signal, which can be observed by examining the maximum values shown in Figure S10.3.

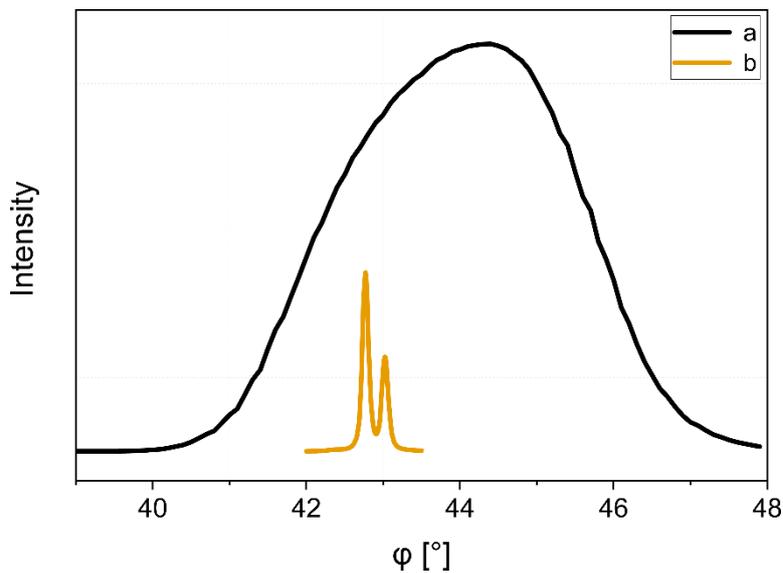

**Figure S10.3.** φ-scans of the (311) reflection of MgO (100) monocrystal. (a) Asymmetrical geometry: 2θ = 74.68°, ω = 12.105°, ψ = 0°. (b) Tilted sample geometry: 2θ = 74.68°, ω = 37.34°, ψ = 25.24°.